\documentclass[twocolumn,showpacs,floatfix,aps]{revtex4}
\usepackage{amssymb,amsmath}
\usepackage[dvips]{graphics,color}
\usepackage{epsfig}\usepackage{float}
\newcommand{\ba}{\begin{eqnarray}}
\newcommand{\ea}{\end{eqnarray}}
\newcommand{\bege}{\begin{equation}}
\newcommand{\enge}{\end{equation}}
\newcommand{\beq}{\begin{eqnarray}}
\newcommand{\benu}{\begin{enumerate}}
\newcommand{\enu}{\end{enumerate}}
\newcommand{\eeq}{\end{eqnarray}}
\newcommand{\pa}{\partial}
\newcommand{\noi}{\noindent}

\begin{document}

\title{Shortcuts in particle production
in a toroidal compactified spacetime }

\author{J. M. Hoff da Silva}
\email{hoff@ift.unesp.br} \affiliation{Instituto de F\'{\i}sica
Te\'orica, Universidade Estadual Paulista, Rua Pamplona 145,
01405-900 S\~ao Paulo, SP, Brazil}
\author{R. da Rocha}
\email{roldao.rocha@ufabc.edu.br, roldao@ifi.unicamp.br} \affiliation{
Centro de Matem\'atica, Computa\c c\~ao e Cogni\c c\~ao,
Universidade Federal do ABC, 09210-170, Santo Andr\'e, SP, Brazil\\and\\
Instituto de F\'{\i}sica ``Gleb Wataghin''\\ Universidade
Estadual de Campinas, Unicamp\\
 13083-970 Campinas, S\~ao Paulo, Brasil}


\begin{abstract}

We investigate the particle production in a toroidal compactified
spacetime due to the expansion of a Friedmann cosmological model
in $\mathbb{R}^{3}\times S^{1}$ outside a $U(1)$ local cosmic
string. The case of a Friedmann spacetime is also investigated
when torsion is incorporated in the connection. We present a
generalization to toroidal compactification of $p$ extra
dimensions, where the topology is given by  $\mathbb{R}^{3}\times
T^{p}$. Some implications are presented and discussed.
% Thoroughout
%the paper we are not solely interested in the dynamics of
%spacetime, but on the physical consequences of the topological
%transformations.
Besides the dynamics of spacetime, we also investigate in details
 the physical consequences of the topological
transformations.

\end{abstract}
\maketitle

\section*{Introduction}

Cosmic strings are a very interesting class of topological
defects, which arise from spontaneous symmetry breaking, and could
be generated by early universe phase transitions \cite{vilenkin,
kib, kib2}. Depending on the broken symmetry, the type of
topological defect may change. In particular, cosmic strings are
related to the  non-triviality of the first homotopy group of
broken symmetry \cite{vilenkin}. At the same time, up to now there
is no consistent physical law that determines the topological
shape of universe. So, a wide range of possibilities can be
experimented, at least, theoretically.

Here we investigate some shortcuts concerned to the particle
production due to an expansion of the universe given by the
following metric $(G=c=\hbar=1)$ \ba
ds^{2}=a(t)(-dt^{2}+dr^{2}+B^{2}r^{2}d\theta^{2}+dz^{2}),
\label{first} \ea where the $z$ direction is compactified in a
circle --- $S^{1}$ topology --- and the string, chosen to be along
this direction, has a finite length $L$. In some sense, it is a
complement of  what was done in \cite{m3} and we shall refer to it
latter.

The line element  in Eq.(\ref{first}) can be derived from
Einstein's equations with source given by the energy-momentum \emph{ansatz} \cite{vilenkin}
 \ba{T^{\alpha}_{\beta}=\mu\, {\rm diag}\,(1,0,0,1)},\label{c2}\ea
where $\mu$ is the string linear density. For instance, this
\emph{ansatz} stands for a straight string and the metric in
Eq.(\ref{first}) is obtained from Eq.(\ref{c2}) in linear gravity.
However, it is possible to show that the form of the metric given
in Eq.(\ref{first}) is still valid for full Einstein's equations,
i.e., if gravity is input in the standard gauge Higgs theory, and
Einstein's equations are solved using the scalar field as source.
In Eq.(\ref{first}) the parameter $B$ is related to $\mu$ by
$B=1-4\mu$, and stands for the well known deficit angle of the
cosmic string. In this paper we consider the GUT scale ($\mu \sim
10^{-6}$).  It is clear that when $B\rightarrow 1$
--- or correspondingly when $\mu\rightarrow 0$ --- the metric in
Eq.(\ref{first}) is
led to the Friedmann standard metric.

The last peculiarity of Eq.(\ref{first}) we want to emphasize is
the $S^{1}$ topology \cite{m3}. Our coordinate system obeys the
equivalence class constraint $(t,r,\theta,z)=(t,r,\theta,z+mL)$,
where $m$ is an integer. After all, Eq.(\ref{first}) represents
the line element outside a straight cosmic string in $r=0$,
parallel to the $z$-axis in a Friedmann cosmological model with
$S^{1}\times \mathbb{R}^{3}$ topology.

In what follows we also analyze the particle production due to an
expansion in such spacetime, including the case where torsion is
present, in the context of Eq.(\ref{first}), in order to solve
Klein-Gordon equation associated with a massless field in a
Riemann-Cartan compactified spacetime. The investigation of this
more general case is important, since torsion gravity theory can
be considered as one of the most natural extensions of General
Relativity \cite{sete,hehl10,h2} and, in the particular
teleparallel connection we will use, the scalar field \emph{does}
couple to torsion \cite{b,b22}. Also, recently there has been
discovered a spin-$1/2$ matter field with mass dimension one,
called ELKO spinor fields \cite{allu,boe2,meu}, which is closely
related to standard scalar field theory presenting
self-interaction term. Such class of spinor fields are prime
candidates for dark matter and inflation \cite{boe2}. Further, a
global torsion field is one possible candidate for the CPT and
Lorentz violation parameters \cite{kos}. Another great motivation
to incorporate torsion in this formalism is that it can be thought
as being a fundamental propagating field, which is characterized
by torsion mass and the values of the coupling between it and
fermions \cite{sete,shap}. It is well known that effective quantum
theory put severe restrictions on the torsion parameters
\cite{shap}, and it could impose significant characteristics
concerning the geometric structures endowing spacetime.

The analysis is generalized to the $\mathbb{R}^{3}\times T^{p}$
topology, where $T^p$ denotes the $p$-torus $S^1\times
S^1\times\cdots\times S^1$. Although the influence of the
 cosmic string background on the particle
production was investigated in, e.g., \cite{g1,sahni,nami} we are
interested here in the way that the unusual topology --- achieved
by $S^1$ and $T^p$ compactification --- can intervenes in this
process.

The article is presented as follows: In Section (\ref{333}) we
analyze the particle production using the solutions of the
Klein-Gordon equation to calculate the Bogoliubov coefficients.
This paradigmatic Section establishes the standard tools for the
subsequent analysis. In Section (\ref{444}) the case where torsion
is incorporated in the connection in investigated in details, and
finally in Section (III) the case without torsion is generalized
for a $p$-dimensional toroidal compactification.

\section{Particle Production}
\label{333} When spacetime is static, even in a curved background
it is easy to define the vacuum state of the system. This is
because it is trivial to find a time-like Killing vector that
generates an one-parameter Lie group of isometries, and so the
vacuum together with  all the Fock space can be defined
\cite{livro}. Let us analyze the particle production of a scalar
massless field in our model due to an expansion of the universe,
i.e., corresponding to a functional form to $a(t)$ in
Eq.(\ref{first}). We should emphasize that this scalar field have
no relation to the Higgs-like scalar field that generates the
string.

The {\it in} and {\it out} states are defined by the static
background (cosmic string) and the expansion shall be implemented
by a special function in the conformal factor. As expected, the
nontrivial topology implies a change in physical events, and the
procedure is very similar to the one found in \cite{g1,sahni}.

The Klein-Gordon equation for the massless scalar field $\phi =
\phi (t,r,\theta ,z)$ is
\begin{equation}\hspace{-0.1cm}\left(
\partial_{t}^{2}+\frac{1}{a}\partial_{t}(a\partial_{t})-\frac{1}{r}\partial_{r}(
r\partial_{r})\right. - \left.
\frac{1}{B^{2}r^{2}}\partial_{\theta}^{2}-\partial_{z}^{2} \right)
\phi =0.\label{part1} \end{equation} Separation of variables gives
$\phi(x^{\mu})=R(r)e^{i\lambda z}e^{i\alpha \theta} T(t)$, where
$T(t)$ and $R(r)$  respectively satisfy the following equations:
\ba \frac{d^{2}T(t)}{dt^{2}}+\frac{\dot
a}{a}\frac{dT(t)}{dt}+w^{2}T(t)=0 \label{part2} \ea and
 \ba \left. \frac{d^{2}R(r)}{dr^{2}}+\frac{1}{r}\frac{dR(r)}{dr} +
\frac{\alpha^{2}R(r)}{B^{2}r^{2}}+(\lambda^{2}-w^{2})R(r)=0\right.
\label{part3}, \ea where $\dot a =da/dt$, and $w$ is a separation
constant. Note that $\alpha$ and $\lambda$ obeys periodical
constraints due to the topology of the cosmic string and $S^{1}$,
respectively. Dirichlet boundary conditions are imposed at
$r=\tilde r$ to keep the produced energy in a limited region
\cite{nami}. Then the solution of Eq.(\ref{part3}) is given by \ba
R(r)=\frac{J_{\nu}(\sqrt{\lambda^{2}-w^{2}}r)}{J_{\nu}(\sqrt{\lambda^{2}-w^{2}}\tilde
r)}-\frac{Y_{\nu}(\sqrt{\lambda^{2}-w^{2}}r)}{Y_{\nu}(\sqrt{\lambda^{2}-w^{2}}\tilde
r)}, \label{part4} \ea where $\nu=\alpha /B$. Besides, the
solution on the string is well defined by imposing an additional
vanishing boundary condition at $r=r_{0}$, i.e., we impose a restriction on the
system. The solution is, then, valid in the range
$r_{0}< r< \tilde r$, and the values of $w$ can be found by the
transcendental equation
 \ba
J_{\nu}(Cr_{0})Y_{\nu}(C\tilde r)-J_{\nu}(C\tilde
r)Y_{\nu}(Cr_{0})=0, \label{part5} \ea arising from the boundary
conditions, where $C=\sqrt{\lambda^{2}-w^{2}}$ and the roots are
labeled by $k \in \mathbb{N}$. To find the Bogoliubov coefficients
we need to know the function $T(t)$ in the \emph{in} and
\emph{out} regions. It is briefly recalled here, and for more
details see \cite{livro}. After a change in time coordinate by
$\tau =\int \frac{dt}{a(t)}$ we have
 \ba
\frac{d^{2}T(\tau)}{d\tau ^{2}}+w^{2}a^{2}(\tau)T(\tau)=0
\label{part6}, \ea leading to a well known case \cite{livro}. Now,
suppose a smooth expansion for this universe, e.g., $a(\tau )=\Big
(\frac{1}{2}[\Omega^{2}+1+(\Omega^{2}-1)\tanh(\rho
\tau)]\Big)^{1/2}$, where $\Omega $ is a constant and $\rho$ gives
the rate of expansion. Finally the solution is given in terms of
hyperbolic  and hypergeometric functions
 \ba T_{\frac{in}{out}}(\tau )&=&\left.
\frac{1}{4\pi ^{2}}(4\pi w_{\frac{in}{out}})^{-1/2} \right.
\nonumber \\ &\hspace{-2mm}\times & \hspace{-2mm} \left.
\exp{(-iw_{+}\tau -i\frac{w_{-}}{\rho}
\ln{[2\cosh{\rho\tau}]})}\right. \nonumber \\&\hspace{-4mm}\times
&\hspace{-4mm}\left. _{2}F_{1}\Big(1+
i\frac{w_{-}}{\rho},i\frac{w_{-}}{\rho}; 1\mp
i\frac{w_{\frac{in}{out}}}{\rho};\frac{1}{2}(1\pm
\tanh{\rho\tau})\Big)\right. \nonumber \label{part7} \ea where
$w_{in}=w$, $w_{out}=w \Omega$ and $w_{\pm}=\frac{w}{2}(\Omega \pm
1)$. The linear transformations of hyperbolic functions give us
the desired relation between \emph{in} and \emph{out} states \ba
\phi ^{in}_{\lambda ,\alpha ,k}=\gamma (\lambda ,\alpha
,k)\phi^{out}_{\lambda ,\alpha ,k}+\beta (\lambda ,\alpha
,k)(\phi^{out}_{\lambda ,\alpha ,k})^{*} \nonumber \label{part8}.
\ea The terms $\gamma$ and $\beta$ above are the so-called
Bogoliubov coefficients given, in our case, by
 \ba \gamma (\lambda
,\alpha ,k)=\Omega \frac{\Gamma[1-iw_{in}/\rho ]\Gamma
(-iw_{out}/\rho )}{\Gamma (-iw_{+}/\rho )\Gamma[1-iw_{+}/\rho
]}\label{part9} \ea and \ba \beta (\lambda ,\alpha ,k)=\Omega
\frac{\Gamma[1-iw_{in}/\rho ]\Gamma (iw_{out}/\rho )}{\Gamma
(iw_{-}/\rho )\Gamma[1+iw_{-}/\rho ]}.\label{part10} \ea According
to the usual theory in curved spaces, the density of created
particles per mode is $|\beta (\lambda ,\alpha ,k)|^{2}$, i.e.,
\ba |\beta (\lambda ,\alpha ,k)|^{2}=\frac{\sinh^{2}{(\pi /\rho
w_{-})}}{\sinh{(\pi /\rho w_{in})}\sinh{(\pi /\rho
w_{out})}}\label{part11}. \ea We should remark that the effects of
nontrivial topology is reflected on the excitation modes. In
Eq.(\ref{part11}) the modes assigned by $\alpha$ must obey the
periodic constraint \ba \alpha =2\pi n. \label{part12} \ea
Besides, the $S^{1}$ compactification in the $z$ coordinate
implements another constraint, this time in $\lambda$, since the
string have a finite length $L$ and these modes have a discrete
spectrum \ba \lambda =\frac{2\pi m}{L} \label{part13} \ea where
$n, m \in \mathbb{Z}$. Note that this is an important result: the
geometry of the source, as well as of the compactified dimension,
modifies the mode of excitation from continue to discrete modes.

\section{The formalism in a Riemann-Cartan spacetime}
\label{444} The Klein-Gordon equation for the massless scalar
field in a Friedmann Riemann-Cartan spacetime, which geometry is
endowed with a connection possessing a non null torsion tensor
$T^{\nu\;\,\lambda}_{\;\;\rho}$,  is given by \cite{b,b22}
\bege \frac{1}{\sqrt{|g|}}\pa_\mu(\sqrt{|g|}g^{\mu\nu}\pa_\nu
\phi) + T^{\nu\;\,\lambda}_{\;\;\nu}\pa_\lambda\phi = 0. \enge The
metric in this case has the same form of Eq.(\ref{first}), but it
must be emphasized that to the connection a new term,
corresponding to the torsion, must be added \cite{b,b22}. It can be shown
that this new solution of Einstein's equation is diffeomorphic to
the one found in Eq.(\ref{first}), and since we are interested in
the topological transformations, they are indeed equivalent.

After separation of variables, only the radial and temporal
functions are modified by torsion effects, and we get \bege
\phi(t,r,\theta,z)=T(t)R(r)\Theta(\theta)Z(z), \enge\noi where \ba
R(r) &=&\left.
\frac{1}{\sqrt{r}}\Big(C_1\,J_{\frac{1}{2}\sqrt{1-4n^2}}(-i\alpha
r)\right.\nonumber\\ &+&\left.
C_2\,Y_{\frac{1}{2}\sqrt{1-4n^2}}(-i\alpha
r)\Big)\right.,\label{0}\ea and in the particular case where the
integration constant $n$ is zero, $R(r)$ reads \bege R(r) = A_1\,
\exp(\alpha r) + A_2\,\exp(-\alpha r). \enge We have to choose in
Eq.(\ref{0}) the term which is regular at the origin, and the
Dirichlet boundary conditions are implemented by the constants
$C_{1}$ [$A_{1}$] and $C_{2}$ [$A_{2}$] for any value of $n$
[$n=0$].

The temporal equation is given by \bege \frac{\pa^2 T(t)}{\pa t^2}
+ 2\dot{a}(t) \sqrt{a(t)} + \frac{\pa T(t)}{\pa t} + a(t)\omega^2
T(t) = 0,\label{XX} \enge\noi where $\omega$ is a separation
constant. Note that there is an extra term arising from torsion
effects. There is not, in general, an analytical function which is
a solution of the equation above. In fact, the set of the
solutions of Eq.(\ref{XX}), that are \emph{not} analytical
explicit functions, is dense. However, in some particular cases of
$a(t)$ the solution is an analytical explicit function. We are
concerned with a particular class of analytical solution $a(t)$
which is a smooth expansion and satisfies $0 < b = \lim_{t\rightarrow +\infty} a(t)$, in
order to the \emph{in} and \emph{out} states to make sense --- again the system is restricted to a box. For
instance, the solution when $a(t) = (t^2 -1)/(t^2+1)$ is given by
\bege T(t) = A\;\cos(f(t)^{1/4}\exp[2f(t)]t + \delta) \enge\noi
where $\delta$ is a phase and $f(t)=\sqrt\frac{ct^2-1}{ct^2+1}$
($c$ = cte.). Also, \bege \Theta(\theta) = D_1\,\cos(\alpha\theta)
+ D_2\,\sin(\alpha\theta) \enge\noi and \bege\label{1} Z(z) =
\exp(i\lambda z), \enge\noi where the condition $Z(z) = Z(z+mL)$
--- in order to accomplish the toroidal compactification --- is
implicit in the form of Eq.(\ref{1}). The Bogoliubov coefficient
is given by $(\phi_{in}, \phi^{*}_{out})$ --- the internal product
between $\phi_{in}, \phi^{*}_{out}$. Its explicit calculation is
beyond the scope of the present article. Nevertheless, it is clear
from the solutions that the geometry also shall transform the
$\lambda$ and $\alpha$ modes in discrete ones.

\section{Compactified extra-dimensional case}
\label{555} Once we have introduced the formalism for one
compactified dimension, the generalization to extra dimensions is
immediate. First of all, we introduce a topological structure with
cylindrical symmetry in the hypersurface $(t,z_{1},\ldots,z_p)$
constant. Our
interest resides in the topological transformations of
spacetime. We start with the
following line element \ba ds^{2}=
a(t)\left(-dt^{2}+dr^{2}+B^{2}r^{2}d\theta^{2}+
\frac{b(t)}{a(t)}dz_{i}dz^{i}\right), \label{extra1} \ea where
$i=1,\ldots,p$. Now the topology is $\mathbb{R}^{3}\times T^{p}$
and we can investigate some shortcuts in particle production from
a more generalized compactification formalism. Note that we are
not considering a cloud of strings in the sense of \cite{cloud}.
Instead, we just deal with an unusual topology without any string
in extra dimensions. Suppose that the universe, after the
expansion analyzed before, passes to another expansion behavior,
now characterized just by the dimensions of the $p$-torus. In
other words, suppose that in a first moment $b(t)=a(t)$, and  all
previous analysis (Sec. (\ref{333})) is still valid. After that, in
$t=t_{0}$, consider an expansion in the extra dimensions and in the
$z$ direction. The Klein-Gordon equation for the field
$\phi=R(r)\Theta(\theta)T(t)Z(z_1,z_2,\ldots,z_p)$ in the geometry
given by Eq.(\ref{extra1})  reads \ba && \left.
\Bigg[\partial_{t}^{2}+\frac{1}{2}\Big( \frac{\dot
b}{b}+\frac{\dot a}{b} \Big) \partial_{t}\Bigg] \phi +
\frac{1}{r}[\partial_{r}(r\partial_{r})]\phi \right. \nonumber
\\&+&\left. \frac{1}{r^{2}B^{2}}\partial_{\theta}^{2}\phi
+\frac{a(t)}{b(t)}\partial_{z_{i}}\partial_{z^{i}}\phi =0. \right.
\label{extra2} \ea Note that if $b(t)=a(t)$ and $p=1$ we recover
the previous situation. If $b(t)=a(t)$ and $p \neq 1$  the
generalization is immediate: we obtain, after separation of
variables in Eq.(\ref{extra2}), the solution
$Z(z_1,z_2,\ldots,z_p)= \prod_{i=1}^p \exp{(\lambda^{i}z_{i})}$.
Of course, there will be another classification for each root of
Eq.(\ref{part5}).

Now consider the case where $a=\Omega$ and $b(t)$ simulate an
expansion until $t=\bar t$. Then, in the range $t_{0}< t < \bar
t$, after the expansion in all universe ($a(t)$) we have
 \ba
\frac{d^{2}T}{dt^{2}}+\frac{1}{2}\frac{\dot b}{b}\frac{dT}{dt}+
\Bigg(\frac{\Omega}{b}\lambda_{i}^{2}+w^{2} \Bigg) T=0
\label{extra3} \ea and \ba
\frac{d^{2}R}{dr^{2}}+\frac{1}{r}\frac{dR}{dr}+\Bigg(
w^{2}-\frac{\alpha^{2}}{r^{2}B^{2}} \Bigg) R=0 \label{extra4}. \ea
The separation of variables is slightly different from previous
cases, and now $t_{0}$ and $\bar t$  give the asymptotic
solutions. The solution for Eq.(\ref{extra4}) is given by \ba
R(r)=C_{1}J_{(\alpha /B)}(wr)+C_{2}Y_{(\alpha
/B)}(wr),\label{extra5} \ea where $C_{1}$ and $C_{2}$ are
constants to be determined, while Eq.(\ref{extra3}) gives, for a
expansion implemented for instance by $\exp({b_{0}t})$, the
solution
 \ba
T(t)=e^{\frac{-b_{0}t}{4}}\Bigg[ \frac{J_{\bar \nu}\Big( \xi
e^{\frac{-b_{0}t}{2}}\Big)}{J_{\bar \nu}\Big( \xi
e^{\frac{-b_{0}\bar t}{2}}\Big)} - \frac{Y_{\bar \nu}\Big( \xi
e^{\frac{-b_{0}t}{2}}\Big)}{Y_{\bar \nu}\Big( \xi
e^{\frac{-b_{0}\bar t}{2}}\Big)} \Bigg] \label{extra6}, \ea where
$\bar \nu =-1/2\sqrt{ \frac{b_{0}^{2}-16w^{2}}{b_{0}^{2}}}$ and
$\xi =\frac{2\lambda_{i}\Omega^{1/2}}{b_{0}}$.

Again, by using some appropriate boundary condition, the Bogoliubov
coefficient that supplies the vacuum excitation can be calculated.
This last model is nothing but a toy model, however it may serve
as a good laboratory to analyze some combined effects of extra
dimensions and an unusual topology.

We emphasize that in this context $b(t)$ is not understood as a
{\it radion} field \cite{grego}, since it has not a continuum
behavior in time.

\section{Concluding remarks and outlooks}

We have investigated how a topology, where one or more dimensions
are compactified, can present signatures in a specific physical
aspect, like particle production. It can be realized that the
topology generated by the cosmic string is codified in the deficit
angle and in the cylindrical symmetry, while the $S^1$
compactification is realized in the transformation from continuous
modes to discrete ones. A general analysis of the Klein-Gordon
massless field in the context of the Riemann-Cartan geometry is
accomplished, formally ELKO Lagrangian is very similar to
 scalar field Lagrangian.
%Eigenspinors of dual helicity ---
ELKO spinor fields obey scalar field-like equations and
 it is shown in \cite{allu,boe2} that they are prime candidates for
inflation and
dark matter. ELKO spinor fields are also more sensitive to the
spacetime torsion \cite{boe2}, and from the formal viewpoint
Section (\ref{444}) gives the essential pre-requisites to
investigate the relationship
 between particle production in a Riemann-Cartan Friedmann spacetime and
the fields
that are prime candidates to describe
 dark matter.

We also investigated, in Section (\ref{555}), the generalization
of extra-dimensional toroidal-compactified models. One interesting
particular case, when the topology is given by $\mathbb{R}^3\times
T^2$, can introduce a Kaluza-Klein theory over AdS spacetime,
since $\mathbb{R}^3\times T^2\simeq (\mathbb{R}^3\times S^1)
\times S^1$, and $\mathbb{R}^3\times S^1$ corresponds to the AdS
topology. In this topology it is also possible to use all the
results arising from Eq.(\ref{extra1}). In this case obviously the
Kaluza-Klein modes are discrete, since we deal with compactified extra
dimensions. The cases presented in Sections (II-III) give rise to
the most direct generalization of our previous analysis and play
an important role in this context, since they have interesting
properties in the particle production processes. Vacuum effects in
the field of multiple cosmic strings \cite{aliev} in compactified
dimensions shall be analyzed in a forthcoming paper.

Some interesting properties concerning scalar fields in the model
shown in Sec.(\ref{333}) were studied in \cite{m3}, like $\langle
\phi^{2}\rangle$ fluctuations, where $\phi$ is a massless scalar
field. In this sense, the complete characterization of the quantum
vacuum requires a deeper understanding about stress tensor
vacuum fluctuations. Many problems appear in the formulation of
such semiclassical system.  Since the basic formalism was
developed in the usual topology like in \cite{l1,propag}, where
the field propagator is well established, and the stress tensor
computed, similar or analogous results were not achieved for the
$\mathbb{R}^3\times S^{1}$ topology, due to the inherent
difficulties of such systems. For instance, this topology gives
origin to ultraviolet divergences, that must be renormalized by
some method. Again, old questions on quantum field theory in
curved spacetimes return: as 1) how to define a quantum state (in
Fock space) in curved global background given by Eq.(\ref{first});
2) how to implement a reasonable cutoff from the physical point of
view and 3) what kind of physically  new and relevant information
we can get from this unusual topology.

Finally, by accomplishing a toroidal compactification, there
naturally arises a maximal number of covariantly constant spinor
fields, since flat tori are the only manifolds with trivial
holonomy. Each one of these spinor fields is related to a
supersymmetry that remains unbroken by the compactification.
Supersymmetric cosmic strings models involving toroidal
compactification are beyond the scope of this paper.

\section{Acknowledgment}
%The authors are very grateful to Prof. for important comments about this
%paper.
%and to Profs.*** for pointing out some missing points
%in a previous version of this paper.
Rold\~ao da Rocha thanks to Funda\c c\~ao de Amparo \`a Pesquisa
do Estado de S\~ao Paulo (FAPESP) (PDJ 05/03071-0) and J. M. Hoff
da Silva thanks to CAPES-Brazil for financial support.

\end{document}